\documentclass[10pt]{article}
\usepackage{amssymb,latexsym,epsfig,cite}

\begin{document}

\title{Nonlinear electrodynamics and black holes}
\author{N. Bret\'on and R. Garc\'{\i}a-Salcedo.\\
Departamento de F\'{\i}sica, Cinvestav-IPN,  Mexico City.}

%\footnote{
%Departamento de F\'{\i}sica, Cinvestav-IPN, \\
%Apdo. Postal 14-740, D.F., M\'exico.\\}}
%e-mail: nora@fis.cinvestav.mx}}

\maketitle

\begin{abstract}
It is addressed the issue of black holes with nonlinear electromagnetic
field, focussing mainly in the Born-Infeld case. The main features of
these systems are described, for instance, geodesics, energy conditions,
thermodynamics. Also it is revised some black hole solutions of
alternative nonlinear electrodynamics and its inconveniences.
\end{abstract}

%\keywords{}

%**********************************************************
\section{Introduction}

Since the begining of the past century, proposals were made of nonlinear
electrodynamics with the aim to cure the singularity of Maxwell's solution
to the field of a point charge at the charge's position. Among the most
succesful is the formulation of Born and Infeld (BI). The proposal by Born
and Infeld \cite{BI} in 1934 has several desirable properties in a
physical theory, for instance finiteness of the electric field and energy
at the charge's position, the freedom of duality rotations, propagation of
discontinuities of the electromagnetic field on single characteristic
surfaces, among others.
 
Born and Infeld inspired in a finiteness principle chose a particular
``nonlinear'' action with a maximum field strength $b$; they solved the
resulting field equations for the static spherically symmetric solution
corresponding to a point charge. The resulting field was very different
from a Coulombian field in the neighborhood of the point charge where the
fields are most intense; they found, in fact, that the field was finite
everywhere, including the location of the point charge.

Solutions to the Einstein equations coupled with Born-Infeld nonlinear
electromagnetic theory were found as soon as 1937 by Hoffman and Infeld
\cite{hoffmann}. Nonlinear electrodynamics coupled to general relativity
was also explored by Peres \cite{Peres}, finding some static spherically
symmetric solutions as well as wavelike solutions.
 
Pellicer and Torrence \cite{PT} faced BI theory coupled with gravitational
field and obtained a class of nonsingular static spherically symmetric
(SSS) solutions, corresponding to a point charge source, that
asymptotically behaves as a Reissner-Nordstrom (RN) solution. RN is the
SSS solution of Einstein-Maxwell equations.
       
During the eighties, Pleba\~nski \cite{Pleban1} and colaborators made an
extensive study of nonlinear electrodynamics (NLED) in general relativity,
in particular they addressed Einstein equations coupled to Born-Infeld
field in metrics of type D in Petrov classification \cite{Pleban2},
\cite{GSP}; amog these solutions there is the nonlinear generalization of
the Reissner-Nordstrom solution. They also investigated the causal
propagation of signals and gave a classification of the characteristic
surfaces (along which discontinuities of the field propagate) for NLED
\cite{Pleban3}.
 
Demianski \cite{Demian} found in 1986 a SSS solution of
Einstein-Born-Infeld equations that is regular at the origin, the so
called {\it EBIon}. This solution has a particle like structure with a
regular center.
 
Critics are the arbitrariness in choosing the NLED Lagrangian, since there
is a lot of Lagrangians that fulfill the physical requirements like the
linear weak field limit. In this sense, BI Lagrangian is exceptional since
it can be considered as an effective Lagrangian for quantum
electrodynamics (QED) in the one-loop approximation, as was shown by
Heisenberg and Euler \cite{HE} and later on by Schwinger \cite{Schwinger}
(see also the review by Delphenich \cite{Delphenich}). If another
justification was needed it came from string theory, where solutions of
the BI equations represent states of D-branes \cite{Gibbons}.

In this contribution we review aspects of NLED black holes according to
the following plan: In Sec. 2 NLED formalism is introduced; Sec. 3 deals
with Einstein-NLED solutions that can be interpreted as black holes; we
focus in the static spherically symmetric (SSS) solutions. Sec. 4
addresses the NLED black hole thermodynamics. In Sec. 5 the relationship
between the ADM and horizon masses of NLED solitons and black holes is
presented. Sec. 6 contains stability aspects of NLED SSS solutions and
comments on recent related research are at the end.

\section{NLED formalism}

NLED Lagrangian depends in nonlinear way of the elecvtromagnetic
invariants.
We assume that the nonlinear electromagnetic field can be described
by a vector potential $A_{\mu}$,

\begin{equation}
F_{\mu \nu}=2A_{[\mu,\nu]},
\end{equation}  
$F_{\mu \nu}$ possesses just one independent invariant and one independent
pseudo-invariant:

\begin{equation}
F=\frac{1}{4}F_{\alpha \beta}F^{\alpha \beta}, \quad
\tilde{G} =\frac{1}{4}F_{\alpha \beta} \tilde{F}^{\alpha \beta},
\end{equation}
where $\tilde{F}^{\alpha \beta}$ denotes the dual of ${F}^{\alpha \beta}$
defined by $\tilde{F}^{\alpha \beta}=(i/2 \sqrt{-g})\epsilon^{\alpha \beta
\gamma \delta}F_{\gamma \delta}$. If NLED Lagrangian is to be invariant
under Lorentz group including reflections, then it must depend on $F$ and
$\tilde{G}^2$. The condition that for weak fields the nonlinear theory
should approximate the linear one must be imposed as well.

\subsection{($F, \tilde{G}$) and ($P, \tilde{Q}$) frameworks}
 
It is convenient to introduce the canonical formalism for the system,
related to Lagrangian by a Legendre transformation. Defining

\begin{equation}
P^{\alpha \beta}= 2 \frac{\partial L}{\partial {F}_{\alpha \beta}}=
\frac{\partial L}{\partial F} {F}^{\alpha \beta}+\frac{\partial
L}{\partial \tilde{G}} \tilde{F}^{\alpha \beta},
\label{mateqFG}
\end{equation}

\begin{equation}
H= \frac{1}{2}P^{\alpha \beta}F_{\alpha \beta}-L(F,G^2),
\end{equation}
we can work with $H(F,\tilde{G}^2)$ or with $H=H(P,\tilde{Q}^2)$ the later
depending on the invariants associated to $P^{\alpha \beta}$,

\begin{equation}
P=\frac{1}{4}P_{\alpha \beta}P^{\alpha \beta}, \quad
\tilde{Q} =\frac{1}{4}P_{\alpha \beta} \tilde{P}^{\alpha \beta}.
\end{equation}
  
The material or constitutive equations (\ref{mateqFG}) express $P_{\alpha
\beta}$ through ${F}^{\alpha \beta}$, $F$ and $\tilde{G}$ ; we shall
assume that they can be inverted to express $F_{\alpha \beta}$ through
${P}^{\alpha \beta}$ and $P, \tilde{Q}$. These equations (also called
Hamilton's equations) are

\begin{equation}
F^{\alpha \beta}= 2 \frac{\partial H}{\partial {P}_{\alpha \beta}}=
\frac{\partial H}{\partial P} {P}^{\alpha \beta}+\frac{\partial
H}{\partial Q} \tilde{P}^{\alpha \beta}.
\end{equation}
 
The $(P, \tilde{Q})$ framework is an alternative form of NLED obtained
from the original one, the $(F, \tilde{G})$ framework, by a Legendre
transformation.  Physically reasonable conditions must be imposed on $H(P,
\tilde{Q})$: $H$ is real; for weak fields nonlinear effects are negligible
and the theory must have the limit of linear electrodynamics, i.e. $H(P,
\tilde{Q})=P+O(P^2, \tilde{Q}^2)$; if parity is to be conserved, then
under transformation of $\tilde{Q}$ into $- \tilde{Q}$, $H$ must stay
invariant, therefore, $H$ depends on $P$ and $\tilde{Q}^2$. Often it is
also required that strong or dominant energy conditions hold.

In this formalism, ${F}^{\alpha \beta}$ is the physically significant
electromagnetic field tensor, while ${P}^{\alpha \beta}$ is a tensor with
no direct physical meaning. While the various nonlinear field theories
succeed in making ${F}^{\alpha \beta}$ well behaved, the tensor
${P}^{\alpha \beta}$ is, in general, divergent at the location of point
sources. The reason is that $\delta$-function sources continue to play a
role in these theories and ${P}^{\alpha \beta}$ {\it absorbs} the
singularities of the sources allowing the field tensor ${F}^{\alpha
\beta}$ be well behaved. 
 
Note that by passing from $(P, \tilde{Q})$ framework to $(F, \tilde{G})$,
we are changing from one NLED theory, characterized by some lagrangian
$L(P, \tilde{Q})$, to another, in general different, corresponding to the
lagrangian $L(F, \tilde{G})$; in the case of Maxwell electrodynamics, both
theories coincide, $L=F=H=P$.

The coupled gravitational and NLED equations are derived from the action

\begin{equation}
S = \int{d^4x \sqrt{-g} \{ R (16\pi)^{-1}-L \} },
\label{NLEDaction}
\end{equation}
where $R$ denotes the scalar curvature, $g:= {\rm det} \vert g_{\mu \nu}
\vert$ and $L$, the electromagnetic part, is assumed to depend in
nonlinear way on the invariants of $P_{\mu \nu}$, in the $(P, \tilde{Q})$
framework, or $L$ depending on the invariants of $F_{\mu \nu}$ in $(F,
\tilde{G})$ scheme. Let us refer here to the former one:

\begin{equation}
L=\frac{1}{2} P^{\mu \nu} F_{\mu \nu} -H(P, \tilde{Q}),
\label{Lagr}
\end{equation}

The energy-momentum tensor and the scalar of curvature are given,
respectively, by

\begin{eqnarray}
4 \pi T_{\mu \nu}&=& H_{,P} P_{\mu \alpha} P^{\alpha}_{ \nu}-g_{\mu
\nu}(2PH_{,P} +\tilde{Q}H_{, \tilde{Q}}-H), \nonumber\\
\label{Tmunu} 
R&=& 8(PH_{,P} +\tilde{Q}H_{, \tilde{Q}}-H),
\end{eqnarray}
where $\partial H/ \partial P= H_{,P}$. Note that the curvature scalar,
$R$, and consequently the trace of $T_{\mu \nu}$, may differ from zero.

The Born-Infeld nonlinear electrodynamics is given by the
structural function $H(P, \tilde{Q})$,

\begin{equation}
H=b^2 \left(1-\sqrt{1-{2P}/{b^2}+ \tilde{Q}^2/{b^4}} \right),
\label{BIH}
\end{equation}
where $b$ is the maximum field strength and the relevant parameter of the
BI theory.

\subsection{NLED energy conditions}
   
Using a timelike vector, $V^{\alpha}$, $V_{\alpha}V^{\alpha} <1$, imposing
local energy density being non-negative amounts to $T_{\mu
\nu}V^{\mu}V^{\nu} \ge 0$; while that the local energy
flow vector be nonspacelike requires that $T_{\alpha \beta}T^{\alpha}_{
\gamma}V^{\beta}V^{\gamma} \le 0$; these are, respectively, the weak
energy condition (WEC) and the dominant energy condition (DEC); both
conditions hold provided

\begin{equation}
H_{,P} >0, \quad (PH_{,P} +\tilde{Q}H_{, \tilde{Q}}-H ) \ge 0.
\end{equation}
The strong energy condition (SEC)
$R_{\mu \nu}V^{\mu}V^{\nu} \ge 0$, using the Einstein equation can be
settled as,

\begin{equation}
R_{\mu \nu}V^{\mu}V^{\nu}=8 \pi(T_{\mu \nu}V^{\mu}V^{\nu} +\frac{T}{2})
\ge 0.
\end{equation} 
 
Note that NLED matter can violate SEC if the trace of the energy-momentum
tensor is negative enough; for instance, BI energy-momentum tensor can
violate SEC. In the case of Maxwell ED ($T=0$) the fulfilment of WEC
implies SEC.

\section{ NLED Black holes}

Hoffmann and Infeld (HI) \cite{hoffmann} solved the Einstein-Born Infeld
coupled equations (EBI) for the spherically symmetric case, imposing the
condition of regularity on the electromagnetic tensor $F_{kl}$ and its
first derivative and the same condition to the metric tensor $g_{kl}$. The
requirement that there be no infinities in $g_{kl}$ forces the
identification of gravitational with electromagnetic mass. For the SSS
line element

\begin{equation}
ds^2= - \psi dt^2+\psi^{-1}dr^2+r^2(d \theta^2 + \sin^2{\theta}d
\phi^2),
\label{sss-metric}
\end{equation}
HI find the solution

\begin{equation} 
\psi = 1- \frac{8 \pi}{r} \int_{0}^{r}{(\sqrt{r^4+1}-r^2)dr}.
\end{equation}
 
However, this form leads to a conical singularity. In the same paper HI
found the regular solution given by the fields $D_{,r}= 1/r$, $E_{,r}=
r^2/(r^4+1)$ and the metric function
  
\begin{equation}
\psi_{HI} = 1- \frac{k}{r}+\frac{8 \pi \gamma}{r}
\int_{0}^{r}{\left(r^2 \ln \left[\frac{r^4}{1+r^4}\right] \right)dr},
\end{equation}
where $k$ is a constant of integration corresponding to $(-2m)$ in
Schwarzschild solution. To have regularity at $r=0$ it must be taken
$k=0$. Therefore we must consider as the gravitational mass the quantity
$4 \pi \int_{0}^{r}r^4T^{t}_{t}dr$, that is the total electromagnetic mass
within a sphere having its center at $r=0$. Thus the regularity condition
shows that electromagnetic and gravitational mass are the same.
    
Pellicer and Torrence (PT) \cite{PT} following the lines of HI
\cite{hoffmann} searched for a EBI SSS solution well behaved at the
origin; they imposed the continuity of $F_{\mu \nu}$ and the Lagrangian in
the neighborhood of the charge; continuity on $F_{\mu \nu}$ and $L(F)$
leads to conditions on $H(F)$. They considered $\tilde{G}=0$, but were not
restricted to BI.

For the SSS line element (\ref{sss-metric}) PT obtained 
 
\begin{equation} 
\psi_{PT}= 1+\frac{d}{r} + \frac{8 \pi}{r} \int_{0}^{r}{H(x)x^2dx},
\label{metrPT}
\end{equation} 
the comparison with Schwarzschild solution gives $d=2m$, the constant $d$
is related to the mass parameter $m$. If one considers the mass as arising
from electromagnetic properties, one can put $m=d=0$. The electromagnetic
field for a charge $e$ is given by

\begin{equation}
F_{\mu \nu}= -\frac{e}{r^2} \frac{\partial H(P,0)}{\partial P}2
\delta^{[0}_{\mu} \delta^{r]}_{\nu},
\label{emfieldPT}
\end{equation} 
also $P=-e^2/4r^4$. Regularity takes place if the following conditions
hold: the integral in (\ref{metrPT}) exists and is finite; the field
(\ref{emfieldPT}) must be finite as $r \to 0$; besides, asymptotically,
for large $r$, $H(P,0) \approx P$. These conditions guarantee the solution
be well behaved. Note that there is still some freedom in chosing the
function $H(P)$ so it can be selected {\it ad hoc} and numerous examples
may
be built.
  
Related to this point we shall address some cases of regular black hole
solutions obtained with ad hoc NLEDs.
     
The existence of two alternative ways, $(P, \tilde{Q})$ and $(F,
\tilde{G})$, of setting NLED, can generate some confusion. Let us refer to
those regular electric black hole solutions that have been obtained
recently \cite{Ayon}.  In deriving the solutions, in the $(P,
\tilde{Q}=0)$ framework, an ad hoc NLED $H(P)$ was chosen to obtain a
regular metric that describes black holes with regular center and RN
asymptotics. The {\it appropriate} $H(P)$ was found imposing regularity
conditions on the metric, in a similar way to the sketched by PT. The so
constructed solutions \cite{Ayon} were using:

\begin{eqnarray}
H(P)&=&P \frac{1-3 \Pi}{(1+ \Pi)^3}+\frac{6}{q^2s} 
\left(  \frac{\Pi}{1+\Pi} \right)^{5/2}, \\
H(P)&=&P / \cosh^2(s \sqrt{\Pi}), \\
H(P)&=&P\frac{\exp(-s \sqrt{\Pi})}{(1+ \Pi)^{5/2}} \left( 1+
\frac{3\Pi}{s}+ \Pi \right),
\end{eqnarray}
where ${\Pi}=\sqrt {-q^2P/2}$ and $s= \mid q \mid /(2m)$; $q=q_e$ and $m$
are, respectively, the parameters identified as charge and mass. The above
presented functions $H(P)$ behave like $P$ at small $P$ and tend to finite
limits as $P \to - \infty$.  The solutions are called regular due to the
finiteness of three invariants: $R, R_{\alpha \beta}R^{\alpha \beta}$ and
$R_{\alpha \beta \gamma \delta}R^{\alpha \beta \gamma \delta}$.

However, there is a clear contradiction with no go theorems that forbid
the existence of SSS solutions with a regular center for whatever $L(F)$
chosen if it is required that for weak fields, $L \sim F$, i.e. if for
weak fields it is demanded a Maxwellian behavior \cite{Bronnikov2},
\cite{Bronnikov}.
       
The explanation of this apparent contradiction is the following: one can
obtain regular solutions choosing ad hoc $H(P, \tilde{Q})$ as was
explained above. These solutions correspond to certain lagrangian $L(F,
\tilde{G})$ that is the one from which one derives the dynamical
equations. That this Lagrangian be well behaved is not guarateed by an
{\it adecuate} selection of the corresponding $H(P, \tilde{Q})$. This is
the case, for instance in \cite{Ayon}: those solutions correspond to
Lagrangians $L(F)$ that suffer branching. So, in spite that solutions are
well behaved in the $P$ framework ($\tilde{G}=0$), they correspond to
different Lagrangians in different regions of space. The branching in
$L(F)$ is due to extrema in the electromagnetic field; associated to
extrema there are singularities that are seen only by photons. These
singularities can either be hidden behind a horizon or naked depending on
the value of one parameter. Additional features of those ``regular"
solutions include, for instance, that the energy density of
electromagnetic field may be negative for some interval of the radial
coordinate; this and more were investigated by Novello et al
\cite{Novello}, \cite{Novello2}.
 
Note, however, that regular solutions with only magnetic charge may exist
\cite{Bronnikov}. Another way to avoid the prohibition of SSS electrically
charged regular structures is to resign of having a regular center, like
the solution by Dymnikova in \cite{Dymnikova} that has a de Sitter center.
It is not excluded neither the possibility of regular solutions that
correspond to Lagrangians depending on both invariants of the
electromagnetic field, $L(F, \tilde{G})$.

\subsection{Type-D solutions with EBI}

BI theory has the property that its equations have an exact $SO(2)$
electric-magnetic duality invariance. In \cite{Pleban2} Type-D solutions
were constructed for BI coupled with Einstein equations allowing for the
freedom of duality rotations. Under the assumption that the natural
tetrads of type-D metrics coincide with the eigenvectors of the
algebraically general nonlinear electromagnetic field and assuming that
the two principal null directions are geodesic and shear-free, all D-type
solutions that are compatible with the scheme of NLED endowed with duality
rotations were found. It turned out that the presence of the acceleration
or rotation parameters prohibits the existence of NLED type-D solutions
with duality rotations. Moreover, by adding axion and dilaton fields, this
invariance may be extended to $SL(2, {\bf R})$ S-duality, relevant to
string theory, which implies a strong-weak coupling duality of such
theories; for dualities in the context of more general nonlinear
electrodynamics see \cite{Gibbons-Rasheed}, \cite{PP},
\cite{Gibbons-Hashimoto}.
     
All type-D solutions of the coupled Einstein and Born-Infeld equations
were determined in \cite{GSP}. There are two classes of them: static and
stationary. The static solutions are exhausted by the BI generalizations
of (i) the Bertotti-Robinson solution, (ii) Reissner-Nordstrom (RN) and
(iii) anti-Reissner-\-Nordstrom. The stationary solutions belong to two
subfamilies: the BI generalization of the NUT $\tilde{B}(+)$ metric which
contains as a limit the BI generalization of RN, and the BI generalization
of the anti-NUT $\tilde{B}(-)$ solution that contains as a limit the
static case (iii).  The NUT solution includes the NUT parameter, $m$,
while anti-NUT changes from $m$ to $(-n)$.  These EBI solutions were
derived including a cosmological constant $\Lambda$ in their
energy-momentum tensor. Since BI generalization of RN solution is of most
interest, we analyze it in detail in what follows.

\subsection{Born-Infeld black hole and EBIon}

The EBI solution for a SSS spacetime (\ref{sss-metric})
is given by the metric function $\psi_{BI}(r)$

\begin{eqnarray}
\psi_{BI}(r)&=& 1-\frac{2m}{r} + \frac{2}{3}b^2 (r^2 - \sqrt{r^4+a^4})+
\frac{4g^2}{3r}G(r), \\ 
G'(r)&=&- (r^4+a^4)^{- \frac{1}{2}},
\label{BImetrfunc}
\end{eqnarray}
where $G'(r)$ denotes the derivative of $G(r)$ with respect to the radial
variable, $m$ is the mass parameter, $g$ is the magnetic (or electric)
charge (both in lenght units), $a^4=g^2/b^2$ and $b$ is the Born-Infeld
parameter given in units of $[\rm{lenght}]^{-1}$. The nonvanishing
components of the electromagnetic field are

\begin{equation}
F_{rt}= g (r^4+ a^4)^{- \frac{1}{2}},
\quad P_{rt}= \frac{g}{r^2}.
\label{FrtBI}
\end{equation}  

The black hole solution given by Gar\-c\'{\i}a-Sa\-la\-zar-Ple\-ba\~nski
\cite{GSP} corresponds to

\begin{equation}
G(r)= \int^{\infty}_{r}{\frac{ds}{\sqrt{s^4+a^4}}}=\frac{1}{2a}  
{\mathbb{F}} \left[ \arccos{ \left( \frac{r^2- a^2}{r^2+a^2} \right)},
\frac{1}{\sqrt{2}} \right],
\label{gPleb}
\end{equation}
where ${\mathbb{F}}$ is the elliptic integral of the first kind. On
the other side, the particle-like solution given by Demianski
\cite{Demian} is

\begin{equation}
G(r)=  \int^{r}_{0}{\frac{-ds}{\sqrt{s^4+a^4}}}=- \frac{1}{2a}
{\mathbb{F}} \left[ \arccos{ \left(
\frac{a^2-r^2}{a^2+r^2}\right)},\frac{1}{\sqrt{2}} \right].
\label{gDem}
\end{equation}
 
The selection of $G(r)$ as in Eq. (\ref{gPleb}) or Eq. (\ref{gDem}) has as
a consequence a different behavior of the solution at $r=0$. The metric
function $\psi_{BI}(r)$ with $G(r)$ given by Eq. (\ref{gPleb}) diverges at
$r \to 0$ (even when $m=0$), corresponding to the black hole solution. The
other one, meaning $\psi_{BI}(r)$ with $G(r)$ given by Eq.  (\ref{gDem}),
is the so called EBIon solution that is finite at the origin (for $m=0$).
The integrals of Eqs. (\ref{gPleb}) and (\ref{gDem}) are related by

\begin{equation}
\int^{\infty}_{r}{\frac{ds}{\sqrt{s^4+a^4}}}+\int^{r}_{0}
{\frac{ds}{\sqrt{s^4+a^4}}}=\frac{1}{a}{\rm K} \left[\frac{1}{2} \right],
\end{equation}
where ${\rm K}[\frac{1}{2}]$ is the complete elliptic integral of the
first kind. In the limit of large distances, $r \to \infty$,
asymptotically the solution approaches Reissner-Nordstrom (RN) solution,
the SSS solution to Einstein-Maxwell equations. Also when the BI parameter
goes to infinity, $b \to \infty$, we recover the linear electromagnetic
(Einstein-Maxwell) RN solution. In the uncharged limit, $b=0$ (or $g=0$),
it is recovered the Schwarzschild black hole. Note that due to the duality
rotation both charges, electric $e$ and magnetic $g$, can be included in
the solution by substituting $g \to \sqrt{e^2+g^2}$.
 
Since this solution is of type D, there is only one nonvanishing Weyl
scalar, $\Psi_2$. For the black hole solution it is

\begin{equation} 
\Psi_2= \frac{m}{r^3}-\frac{g^2r^3}{6} \partial_{,rr} \left( \frac{1}{r^2}
\int_{r}^{\infty} \frac{ds}{s^2+ \sqrt{s^4+a^4}} \right),
\end{equation}

\begin{figure}
\centering \epsfig{file=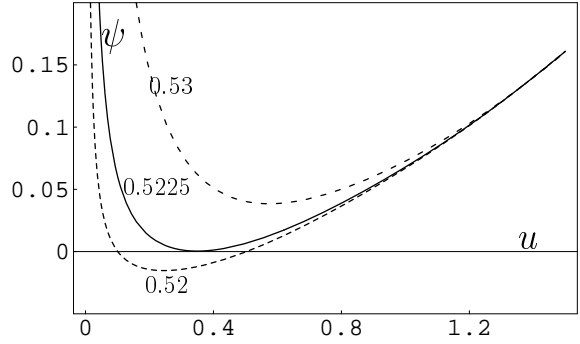, width=8cm} 
\caption{ 
It is shown the behavior of the BI metric function $\psi$ in terms of
$u=r/m$; the label on each curve corresponds to the value of $bm$. Note
that the position of the horizon, $\psi(r_{\Delta})=0$, depends on this
parameter: the horizon shrinks as $bm$ is greater.
} 
\label{psi} 
\end{figure}

The invariants depend on $\Psi_2^2$, then at $r=0$ there is a singularity
of order $1/r^6$, coming from the gravitational mass term, alike the
Schwarzschild and RN singularities; furthermore, the second term also
diverges at $r=0$. There are also zeros in the metric function
$\psi_{BI}(r)$ that are coordinate singularities which can be removed
using analytical extensions \cite{Graves}. The zeros of $\psi_{BI}(r)$ can
be localized numerically and might be one, two or none, depending on the
relative values of the parameters $g, m, b$; it is illustrated in
Fig.\ref{psi}. These parameters determine the position and size of the
horizon as well \cite{Breton1}. Note that distinct $b$ corresponds to
different NLED theories.

\subsection{Trajectories of test particles in BI black hole}
     
Since the SSS spacetime possesses two Killing vectors, $\partial_t$ and
$\partial_{\phi}$, the test particle conserves two motion quantities: its
energy $E$ and its angular momentum $l$. Moreover, if we restrict
ourselves to the equatorial plane ($\theta= \pi/2$), the timelike and null
geodesics for the EBI spacetimes can be reduced to the problem of ordinary
one-dimensional motion in an effective potential $U_{eff}$, alike in the
Schwarzschild and RN cases.

\subsubsection{Massive particles}
 
Trajectories of massive particles are determined by the Lorentz equation.
For a test particle of charge $\epsilon$ and mass $\mu$ it is

\begin{equation}
\frac{d^2x^{\nu}}{d \tau^2}+ \Gamma_{\alpha \beta}^{\nu}
\frac{dx^{\alpha}}{d \tau}\frac{dx^{\beta}}{d \tau}= -
\frac{\epsilon}{\mu}F_{\sigma}^{\nu}
dx^{\sigma}d \tau,
\end{equation}
where $\tau$ is the affine parameter along the trajectorie.  Using the two
conserved motion quantities: energy $E$ and angular momentum $l$, the
geodesic for $t$ can be integrated once, obtaining the first derivative
with respect to the proper time $\tau$,

\begin{equation}
\dot{t} \psi_{BI}=E+ \frac{\epsilon g}{\mu} \sqrt{\frac{b}{4g}}
{\mathbb{F}} \left[
\arccos(\frac{r^2-g/b}{r^2+g/b}), \frac{1}{\sqrt{2}}\right].
\end{equation} 
From the line element for timelike geodesics we have,

\begin{equation}
1= \psi \dot{t}^2- \psi^{-1} \dot{r}^2-r^2 \dot{\phi}^2
\end{equation} 
substituting $l=g_{\phi \phi}\dot{\phi}$ it is obtained

\begin{equation}
\dot{r}^2+ \psi \left( \frac{l^2}{r^2} +1 \right)-
\left[E+ \frac{\epsilon g}{\mu} \sqrt{\frac{b}{4g}}{\mathbb{F}}
\left[
\arccos(\frac{r^2-g/b}{r^2+g/b}), \frac{1}{\sqrt{2}}\right] 
\right]^2=0.
\end{equation} 
Comparison with $\frac{1}{2}\dot{r}^2+U_{eff}(E, l, r)=0$, gives the
effective potential that a charged test particle feels. The shape of the
effective potential shows attractive regions with stable equillibrium
positions. The equilibrium positions are of lower energy for greater
angular momentum of the particles. If test particles reach the singularity
or not depends on the value of the maximum fiel strength $b$.

\subsubsection{Light trajectories in NLED}

\begin{figure}\centering
\epsfig{file=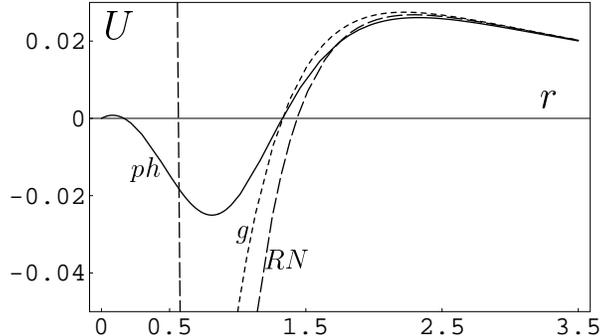, width=8cm}
\caption{
Effective potentials $U_{eff}$ are shown for massless particles in
Reissner-Nordstrom (RN), null geodesics (g), BI photons (ph). The
constants in the plot are: $g=0.6$, $b=0.75$ and angular momentum
$l=1.$
}
\label{U_eff}
\end{figure}
  
Discontinuities of the fields propagate obeying the equation of the
characteristic surfaces, that in ordinary optics is the so-called {\it
eikonal equation}. Locally these surfaces are normal to the light rays
trajectories. In General Relativity when a linear Maxwellian field is
present, its characteristics do coincide with the Einsteinian cone; but
NLED photons do not propagate along null geodesics of the background
geometry. Instead, they propagate along null geodesics of an effective
geometry which depends on the nonlinear electromagnetic field.  In a
curved spacetime the characteristic surfaces equation is
\cite{Pleban3}, \cite{Boillat} 

\begin{equation}
g^{\mu \nu}S_{,\mu}S_{\nu}=0.
\label{null-geod}
\end{equation}

However, if in the spacetime it is present a nonlinear electromagnetic
field, the corresponding equation is

\begin{equation}
(g^{\mu \nu}+ \frac{4 \pi}{b^2}T^{\mu \nu}_{NLED})S_{,\mu}S_{\nu}=
\gamma^{\mu \nu}S_{,\mu}S_{\nu}=0,
\label{phot-traj}
\end{equation} 
where $T^{\mu \nu}_{NLED}$ is the electromagnetic energy momentum density;
remind that the linear limit is obtained when $b \to \infty$, hence in
linear limit light trajectory, that is normal to characteristic surfaces,
occurs on null geodesics.  Hence, as far as $b$ is finite, there is a
distinction between the propagation of gravitational discontinuities
(gravitons, abusing of languaje) along Eq. (\ref{null-geod}) and the
propagation of electromagnetic discontinuities (photons) ruled by Eq.
(\ref{phot-traj}). For BI black holes, actually both trajectories converge
at the horizon.
 
Using the two constants of motion, $E=- \psi \dot t$, $l=\dot \phi r^2$
and the line element for null geodesics, we obtain the derivative of $r$
respect to an affine parameter,

\begin{equation}
\dot r  = \sqrt{E^2- \frac{\psi_{BI} l^2}{r^2}}.
\end{equation}

The trajectories for light rays, however, are given by an effective
geometry, considering the same constants of motion $E$ and $l$:

\begin{equation}
\dot r_{ph}  = \sqrt{E^2- \frac{\psi_{BI} l^2}{r^2} \left( 1+
\frac{a^4}{r^4}
\right)^{-1}}.
\end{equation}
     
There is a correction factor in the last formula, $(1+a^4/r^4)^{-1}$, due
to the effective geometry; it vanishes if $a=0$ ($b \to \infty$) or for
photons with $l=0$. Remind that usually it is considered $\hbar E$ and
$\hbar l$ as the total energy and angular momentum of photon,
respectively. The corresponding effective potentials are shown in Fig.
\ref{U_eff}.
  
From the above expressions it is easy to check that $(dr/dt)_{photon}<
(dr/dt)_{grav}$, i. e. light travels slower that gravitational waves due
to nonlinear effects, such as photon-photon interaction, for instance.  
This effect can also be described as if photons ruled by Maxwell ED were
propagating inside a dielectric medium, with dielectric ``constant''
depending in nonlinear way of the fields. The possibility is not excluded
of superluminal signals for other NLED \cite{Novello}.

\section{NLED black hole thermodynamics}
   
In a general context the zeroth and first laws of black hole mechanics
(BHM) refer to equilibrium situations and small departures therefrom.
First law of BHM is an identity relating the changes in mass, angular
momentum and horizon area of a stationary black hole when it is perturbed.
The variation applies for perturbations from one stationary axisymmetric
solution of Einstein equations to another; moreover, it has been shown
that the validity of this law depends only on very general properties of
the field equations \cite{Wald}. For the horizon mass $M_{\Delta}$ the
first law, when static spherically symmetric solutions are considered, is

\begin{equation}
\delta M_{\Delta}= \frac{\kappa}{8 \pi} \delta{a_{\Delta}}
+ \Phi_{\Delta} \delta Q_{\Delta},
\label{first}   
\end{equation}
where $\kappa$ is the surface gravity at the horizon, $a$ is the area of
the horizon, $Q$ is the electric charge and $\Phi$ is the 
electric potential; the subindex $\Delta$ indicates that the quantity is
evaluated at the horizon of the black hole.

On the other side, the total mass is given by the Smarr's formula

\begin{equation}
M_{\Delta}= \frac{\kappa a_{\Delta}}{4 \pi} +
\Phi_{\Delta} Q_{\Delta} .
\label{smarr}
\end{equation}
 
In the case of Einstein-Maxwell theory, it is possible to deduce one, Eq.  
(\ref{first}), directly from the other, Eq. (\ref{smarr}), using the
homogeneity of the mass as a function of $\sqrt{a}$ and $Q$. In the work
by Ashtekar, Corichi and Sudarsky \cite{ACS} the first law of BHM, for
quantities defined only at the horizon, arises naturally as part of the
requirements for a consistent Hamiltonian formulation. 
  
Work on NLED black hole thermodynamics includes the derivation of the
first law of black hole physics for some nonlinear matter models
\cite{Heusler}. Rasheed \cite{Rasheed} studied the zeroth and first laws
of black hole mechanics in the context of non-linear electrodynamics
coupled to gravity. In this case, the zeroth law, which states that the
surface gravity of a stationary black hole is constant over the event
horizon, is shown to hold even if the Dominant Energy Condition is
violated. In NLED one no longer has homogeneity of the mass function and a
priori one has no reason to expect that neither the first law or Smarr
formula hold. Rasheed found that the usual first law (the general mass
variation formula) holds true for the case of non-linear electrodynamics
but the formula for the total mass (Smarr's formula) does not. 

However, we can propose the form that must have a Smarr-type formula for
the horizon mass in order to be consistent with the variations expressed
by the first law of BHM that indeed holds,

\begin{equation}
M_{\Delta}= \frac{\kappa a_{\Delta}}{4 \pi} +
\Phi_{\Delta} Q_{\Delta}  + V(a_{\Delta}, Q_{\Delta}, P_{\Delta}),
\label{hormass0}
\end{equation}
where $V$ is a so far undetermined potential that depends on the horizon
parameters, $a_{\Delta}, Q_{\Delta}, P_{\Delta}$ and also on the coupling
constants of the theory. In the variational principle this term plays no
role, however in the Hamiltonian description it becomes essential.
 
Note that in the first law, Eq. (\ref{first}), only variations of the
electric charge are involved, and not variations of the magnetic charge,
$P_{\Delta}$. On the other hand, the horizon mass, Eq. (\ref{hormass0})
might depend on $P_{\Delta}$ through $V$.

The equations to determine the potential $V(a_{\Delta},
Q_{\Delta}, P_{\Delta})$ arise from the condition that the first law
holds and demanding consistency between Eq.(\ref{first}) and the
variations of Eq.(\ref{hormass0}), these are \cite{ulises},

\begin{eqnarray}
a_{\Delta} \frac{\partial \psi}{\partial a_{\Delta}}+
8 \pi r_{\Delta} Q_{\Delta}  \frac{\partial \Phi}{\partial
a_{\Delta}} + 8 \pi r_{\Delta}  \frac{\partial V}{\partial   
a_{\Delta}}&&=0,  \nonumber\\
\frac{r_{\Delta}}{2} \frac{\partial \psi}{\partial Q_{\Delta}}+
Q_{\Delta}  \frac{\partial \Phi}{\partial Q_{\Delta}}
+ \frac{\partial V}{\partial Q_{\Delta}}&&=0,
\label{Veqs}
\end{eqnarray}
where $\psi$ is the metric function in a SSS line element
(\ref{sss-metric}), $\psi= 1-2m'(r)$, $a_{\Delta}=4 \pi r_{\Delta}^2$;
$r_{\Delta}$ is the radius of the horizon.
  
The condition of consistency determines the set of parameters that can
vary independently; in this case, the magnetic charge becomes a function
of the area and electric charge, $P_{\Delta}= P_{\Delta}(r_{\Delta},
Q_{\Delta})$. To illustrate the point, in what follows we shall determine
the horizon mass from a Smarr type formula in agreement with the first law
of BHM for the Bardeen black hole.
 
\subsection{ Smarr's formula for Bardeen black hole}
  
The Bardeen model was proposed some years ago as a regular black hole,
however, only recently it has been shown \cite{ABGBardeen} that it is an
exact solution of the Einstein equations coupled to a kind of nonlinear
electrodynamics characterized by the Lagrangian

\begin{equation}
{\cal L}(F)= \frac{2}{2sg^2}(\frac{2g^2F}{1+ \sqrt{2g^2F}})^{5/2},       
\end{equation}
where $g$ and $F$ are the magnetic charge and electromagnetic invariant,
respectively and $s=g/m$.  The corresponding energy momentum tensor
fulfills the weak energy condition and is regular everywhere. For a SSS
space (\ref{sss-metric}), the corresponding metric function is given by

\begin{equation}
\psi_{B}= 1- \frac{2m(r)}{r}= 1- \frac{2mr^2}{(r^2+g^2)^{3/2}},
\label{psi-bardeen}
\end{equation}

This solution is a self-gravitating magnetic monopole with charge $g$. The
solution is regular everywhere, although the invariants of the
electromagnetic field exhibit the usual singular behaviour of magnetic
monopoles, $F= g^2/2r^4$. In the asymptotic behaviour of the solution the
constant $g$ vanishes as $1/r^3$, and not as a Coulombian term ($1/r^2$),
that allows to interpret the constant $g$ as a magnetic charge. The
Bardeen solution does not involve electric charge, then the horizon mass
depends only on the area of the horizon,

\begin{equation}
M_{\Delta}= \frac{1}{8 \pi} \int{\kappa da}= \int{(1-m')dr},
\label{hormass2}
\end{equation}   
the condition that the horizon mass be positive, from Eq.
(\ref{hormass2}), gives that $m(r) \le r$; it also guarantees that
$\psi_{B} \ge 0$. Using the expression for $\psi_{B}$ it amounts to
$(r^2+g^2)^3 \ge 4m^2r^4$. In this case when $g^2=\frac{16}{27}m^2$ the
two horizons that could be present shrink into a single one, being this
value of $g$ the corresponding to the extreme black hole; for $g^2<
\frac{16}{27}m^2$ there exist both inner and event horizon. The potential
$V$ for the Smarr-type formula, Eq. (\ref{hormass0}), for the Bardeen
black hole turns out to be, undetermined until an integration constant
which we have put zero,

\begin{equation}
V= mr^3 \frac{2g^2-r^2}{(g^2+r^2)^{\frac{3}{2}}},
\end{equation}

Substituting $V$ in the Smarr-type formula one obtains the
horizon mass
\begin{equation}
M_{\Delta}= \frac{r}{2}-\frac{mr^3}{(r^2+g^2)^{\frac{3}{2}}}
\label{hormass3}
\end{equation}
 
\begin{figure}\centering
\epsfig{file=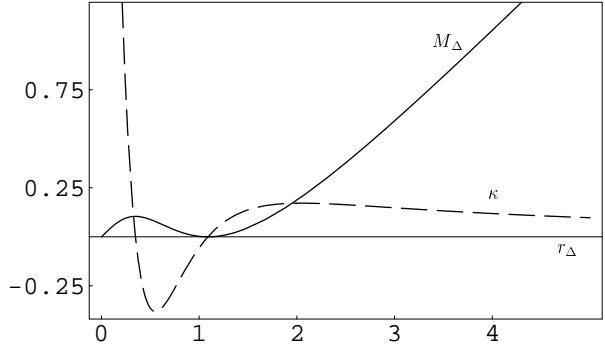, width=8cm}
\caption{
Horizon mass $M_{\Delta}$ and surface gravity $\kappa$, as functions of  
the horizon radious, for the extreme Bardeen black hole, in this case the
magnetic charge has the value $g^2=16m^2/27$.}
\label{Bbhextreme}
\end{figure}

This value for the horizon mass coincides with the one determined by
integrating the first law, Eq. (\ref{hormass2}). Note that the horizon
mass of the Bardeen black hole involved dependence only on the horizon
area, since the magnetic charge is not considered as a varying parameter
of the horizon. In this case there is a full agreement between the horizon
mass calculated with the first law of BHM and when it is determined by
adding the apropriate potential to a Smarr-type formula; it is shown in
Fig. \ref{Bbhextreme}.

The horizon mass, calculated with a Smarr type formula that is consistent
with the fist law of BHM, applies only to the magnetic sector of NLED
solutions. If the variation of electric charge is taken into account in
the potential $V$ of the Smarr formula, the mentioned consistency does not
longer hold.  The potential $V$ determined in agreement with the first law
of BHM can not give the appropriate dependence for the terms corresponding
to electric charge; Eqs. (\ref{Veqs}) do not describe confidently the
potential $V$ in general situations where nonlinear electromagnetic fields
are present. Therefore, when the dependence of $V$ on the electromagnetic
field is of nonlinear nature Eqs.(\ref{Veqs}) are useless to determine
$V$.

\section{Isolated horizon framework and mass relation}

Remarkable properties of nonlinear electrodynamics black holes arise in
the context of the isolated horizon formalism, recently put forward by
Ashtekar {\it et al} \cite{Ashtekar}.  In this approach it is pointed out
the unsatisfactory (uncomplete) description of a black hole given by
concepts such as ADM mass and event horizon, for instance, specially if 
one is dealing with hairy black holes. To remedy this uncompleteness,
Ashtekar {\it et al} have proposed alternatively the isolated horizon   
formalism, that furnishes a more complete description of what happens in
the neighborhood of the horizon of a hairy black hole.

In the isolated horizon formalism one considers spacetimes with an
interior boundary, which satisfy quasi-local boundary conditions that
insure that the horizon remains isolated. The boundary conditions imply
that quasi-local charges can be defined at the horizon, which remain
constant in time. In particular one can define a horizon mass, a horizon
electric charge and a horizon magnetic charge.

Moreover, Ashtekar-Corichi-Sudarsky (ACS) conjecture about the
relationship between the colored black holes and their solitonic analogs
\cite{ACS}: the Arnowitt-Deser-Misner (ADM) mass contains two
contributions, one attributed to the black hole horizon and the other to
the outside hair, captured by the solitonic residue.  In this model, the
hairy black hole can be regarded as a bound state of an ordinary black
hole and a soliton.  The proposed formula relating the horizon mass and
the ADM mass of the colored black hole solution with the ADM mass of the
soliton solution of the corresponding theory is

%%%%%%%%%%%%%%%%%%%%%%%%%%%%%%%%%%%%%%%%%%%%%%%%%%%%%%%%%%%%%%%
 
\begin{equation}
M^{(n)}_{sol}=M^{(n)}_{ADM} -M^{(n)}_{\Delta},
\label{massrel}
\end{equation}

\noindent where the superscript $n$ indicates the colored version of the
hole; in the papers of Ashtekar {\it et al} this $n$ refers to the
Yang-Mills hair, labeled by this parameter, corresponding to $n=0$ the
Schwarszchild limit (absence of YM charge). This relation has been
proved numerically to work for the Einstein-Yang-Mills (EYM) black hole.

For the EBI black hole the location and size of the horizon depends on the
parameter $bq$, so $b$ and $q$ are not independent parameters;  however,
at infinity it is undistinguisable from a RN black hole characterized only
by its charge $q$ and mass $m$.  Provided that in the EBI theory there
exist both exact solutions: the black hole and the soliton like solution,
in spite that the EBI black hole is not a coloured one, we shall probe it
with ACS model, considering $b$ as a free parameter, for a fixed charge.
Then for the case studied here the $n$ version shall correspond to the
distinct black holes labeled by distinct (continuous) BI parameter, $b$.
  
It turns out that the EBI black hole and the corresponding EBIon solution
fulfill the relation between the masses as well as most of the properties
of the model as for the colored black hole \cite{Breton2}. For the EBI
solution the horizon and ADM masses as functions of the horizon radius
$r_{\Delta}$ are given, respectively, by

\begin{equation}
M^{(b)}_{\Delta}(r_{\Delta})= \frac{r_{\Delta}}{2}+\frac{b^2
r_{\Delta}}{3}(r_{\Delta}^2-\sqrt{r_{\Delta}^4+a^4})-\frac{2g^2}{3}
\int^{r_{\Delta}}_{0}{\frac{ds}{\sqrt{a^4+s^4}}},
\label{Mhor}
\end{equation} 

\begin{equation}
M^{(b)}_{ADM}(r_{\Delta})= \frac{r_{\Delta}}{2}+\frac{b^2
r_{\Delta}}{3}(r_{{\Delta}}^2-\sqrt{r_{{\Delta}}^4+a^4})+\frac{2g^2}{3}
\int^{\infty}_{r_{\Delta}}{\frac{ds}{\sqrt{a^4+s^4}}}.
\label{Madm}
\end{equation}

\begin{figure}\centering
\epsfig{file=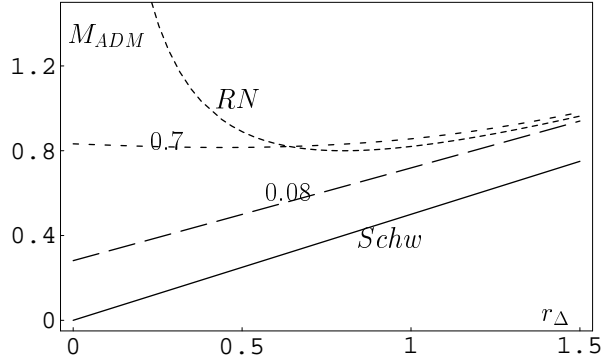, width=8cm}
\caption{
It is shown the ADM mass as function of the horizon radius $r_{\Delta}$,
for the Schwarzschild (Schw), Reissner-Nordstrom (RN) and for BI black
holes with BI parameters $b=0.7$ and $b=0.08$.
}
\label{M_ADM}
\end{figure} 
  
\begin{figure}\centering
\epsfig{file=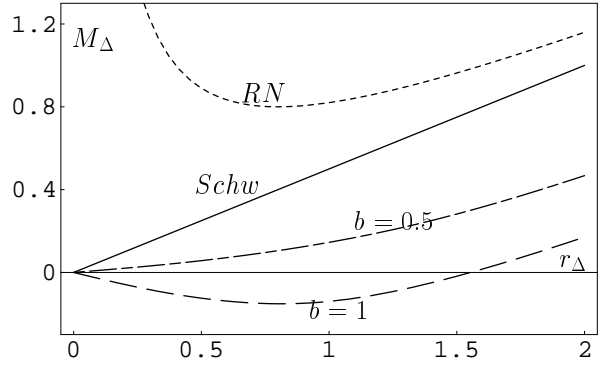, width=8cm} 
\caption{The horizon masses are shown for the Schwarzschild (Schw),
Reissner-Nordstrom (RN) and the BI black    
holes with BI parameters $b=0.5$ and $b=1$.
Note that $M^{RN}_{\Delta} >M^{Schw}_{\Delta}>M^{BI}_{\Delta}$.
}
\label{M_hor}
\end{figure}

In Figs. \ref{M_ADM} and \ref{M_hor} are displayed $M^{(BI)}_{ADM}$ and
$M^{BI}_{\Delta}$ in comparison with the Schwarzschild and
Reissner-Norsdtrom cases. The mass of the soliton can be obtained by
letting $r_{\Delta} \to 0$ in the ADM mass, Eq. (\ref{Madm}), obtaining
$M_{sol}^{(b)}=2g \sqrt{gb}{\rm K}[\frac{1}{2}]/3$. From these expressions
one can trivially check that they satisfy Eq. (\ref{massrel}). 
 
Other predictions of ACS model that are fulfilled by BI black holes and
its soliton counterpart are: BI horizon mass is less than Schwarzschild
horizon mass (see Fig. \ref{M_hor}); horizon masses satisfy the inequality
$M^{RN}_{\Delta} >M^{Schw}_{\Delta}>M^{BI}_{\Delta}$; for all $b$ and all
$r_{\Delta}$, the surface gravity of the BI black hole is less than the
one for Schwarzschild; $M^{BI}_{\Delta}$ and $\kappa_{BI}$ as functions of
$r_{\Delta}$ are monotonically decreasing functions of $b$. However, EBI
solutions do not fulfil that $M^{BI}_{\Delta}$ as function of $r_{\Delta}$
increases monotonically for all values of $b$ (see Fig. \ref{M_hor} for
$b=1$)
  
Since most of the ACS features are fulfilled, we can say that the static
sector of the EBI theory is described by the heuristic model for the
colored black holes proposed by Ashtekar {\it et al} when the BI parameter
$b$ varies keeping the charge fixed.

\section{Stability of NLED black holes}
  
Stability properties in self-gravitating nonlinear electrodynamics were
studied by Moreno and Sarbach \cite{M-S}.  They derived sufficient
conditions for linear stability with respect to arbitrary linear
fluctuations in the metric and in the gauge potential, $\delta g_{\mu
\nu}$ and $\delta A_{\mu \nu}$, respectively; the conditions were obtained
in the form of inequalities to be fulfilled by the nonlinear
electromagnetic Lagrangian $L(F)$ and its derivatives. The application of
this criterion is restricted to static, spherically symmetric solutions of
NLED coupled to gravity, that are purely electric or purely magnetic ($
\tilde{G}=0$). For these systems a gauge invariant perturbation formalism
was used obtaining that linear fluctuations around a SSS purely electric
(or purely magnetic) solution are governed by a wavelike equation with
symmetric potential. The stability conditions are translated into some
requirements on the Lagrangian and its derivatives, $L(F),L_{F}, L_{FF}$;
in terms of the variable $y= \sqrt{2g^2F}$ these conditions are

\begin{equation}                
L(y) >0, \quad L(y)_{,y} >0, \quad L(y)_{,yy} >0 .
\label{stabcond1}
\end{equation}

Besides, there are more inequalities to be fulfilled, that arise from  
the pulsation equations in the even-parity sector

\begin{equation}
f(y) \equiv yL_{,yy}/L_{,y}>0, \quad
f(y)N(y)< 3.
\label{stabcond2}
\end{equation}    
where $N(y)$ is the metric function in  the SSS line element
(\ref{sss-metric}).
 
We shall apply this criterion to test the stability of the purely magnetic
or purely electric EBI particle-like and black hole solutions
\cite{Breton3}. In the former case the boundary point is the origin,
$r=0$, while for the black hole case the conditions must be held in the
domain of outer communication (DOC), i.e. positions outside the horizon,
$r> r_{\Delta}$, $r_{\Delta}$ being the radius of the horizon of the black
hole.
   
The BI Lagrangian fulfills the stability conditions; in terms of the
variable $y$, the BI Lagrangian, with $\tilde{G}=0$, is given by

\begin{equation}
L(y)=b^2[\sqrt{1+\frac{y^2}{b^2g^2}}-1]>0,
\end{equation}
and the rest of the inequalities (\ref{stabcond2}) read as:

\begin{eqnarray}
L_{,y}&=&\frac{y}{g^2}(1+\frac{y^2}{b^2g^2})^{-\frac{1}{2}}>0,
\nonumber\\
L_{,yy}&=&\frac{1}{g^2}(1+\frac{y^2}{b^2g^2})^{-\frac{3}{2}}>0,\nonumber\\
f(y)&=& y\frac{L_{,yy}}{L_{,y}} =(1+\frac{y^2}{b^2g^2})^{-1}>0,
\label{BIineq}                  
\end{eqnarray}
 
Conditions (\ref{BIineq}) are fulfiled in all the range of $y$. Moreover,
$f(y)$ is monotonically decreasing with $f(y=0)=1$, $0<f(y)  \le 1$; then
the last stability condition $f(y)N(y) < 3$ reduces to prove that $N(y)=
\psi_{BI}(y)< 3$; for the black hole it must be fulfilled in DOC
($r>r_{\Delta}$), while for the particle-like solution the domain to be
considered is $0 \le r < \infty$.
  
In the black hole case, the metric function $\psi_{BI}(r)$ has a minimum
in the extreme case ($g=m$) for $bm=0.5224$ at $r_{\Delta}=0.346 m$; DOC
is considered for distances larger than the radius of the horizon, $r>
r_{\Delta}=0.346 m$. In terms of $y$, considering that $F=g^2/2r^4$ then
$y=g^2/r^2$, the metric function $N(y)=\psi_{BI}(y)$ is

\begin{equation}
\psi_{BI}(y)=1-\frac{2m \sqrt{y}}{g}+
\frac{2b^2g^2}{3y}[1-\sqrt{1+\frac{y^2}{b^2g^2}}]+
\frac{2 \sqrt{gby}}{3}{\mathbb{F}} \left[
\arccos({\frac{gb-y}{gb+y}}),
\frac{1}{\sqrt{2}} \right],  
\end{equation}
   
In the range $0<y<y_{\Delta} =8.35$ it turns out that $0< \psi_{BI}(y) \le
1$ with $\psi_{BI}(0)=1$ therefore, $0<N(y)=\psi_{BI}(y)<1<3$, fulfilling
the last inequality required as sufficient conditions for linear stability
of the EBI black hole.

For the particle-like solution of the EBI equations, the metric
function $\psi_{BI}(r)$ in terms of $y$ is

\begin{equation}
\psi_{BI}(y)=1-\frac{2m \sqrt{y}}{g}+
\frac{2b^2g^2}{3y}[1-\sqrt{1+\frac{y^2}{b^2g^2}}]-
\frac{2 \sqrt{gby}}{3}{\mathbb{F}} \left[
\arccos({\frac{y-gb}{gb+y}}),
\frac{1}{\sqrt{2}} \right],
\end{equation}

The last stability condition $N=\psi < 3$ in fact occurs since $N(y=0)=1$
and the function is monotonically decreasing, having $N(y) \le 1$, for $bg
\ne 0$; the finiteness in the origin of $N(r)$ is valid when $m=0$.
Therefore, as far as this analysis proves, the EBI solutions, both black
hole and particle-like one, are stable.
 
On the other side, stability of self-gravitating structures can also be
approached from the isolated horizon framework: the ACS heuristic model
explains the instability of the coloured black holes in terms of the
instability of the solitons. The bound state of a bare black hole and a
soliton is going to be unstable if the total energy in the initial bound
state $E_{initial}= M_{\Delta}^{(n)}(r_{\Delta}^{initial})+M_{sol}^{(n)}$
exceeds the energy in the final black hole,
$E_{final}=M_{\Delta}^{(0)}(r_{\Delta}^{final})$.

The available energy can be expressed (for a fixed
$r_{\Delta}^{initial}$) as

\begin{equation}
E^{(n)}_{avail}=M_{ADM}^{(n)}-M_{ADM}^{(0)},
\end{equation}
where the superscript $n$ indicates the colored version of the hole and
$M_{ADM}^{(0)}$ is the ADM mass of the bare (Schwarzschild) black hole.
The positivity of the difference between the ADM mass and the horizon
mass, $M_{ADM}-M_{\Delta}=E >0$, indicates that there exists an energy $E$
available to be radiated. For static black holes this result can be
interpreted as a potential unstability, i.e. a slightly perturbation in
the initial data will lead the solution to decay to a Schwarzschild black
hole.

The difference between the ADM mass of the BI black hole and
the bare black hole using (\ref{Madm}), turns out to be

\begin{equation}
M^{(b)}_{ADM}(r_{\Delta}) - \frac{r_{\Delta}}{2}=
\frac{b^2}{3}
\left(r_{{\Delta}}^3(1-\sqrt{1+\frac{a^4}{r_{\Delta}^4}})+
a^3{\mathbb{F}} \left[ \arccos{ \left( \frac{r_{\Delta}^2-
a^2}{r_{\Delta}^2+a^2} \right)},
\frac{1}{\sqrt{2}} \right] \right),
\end{equation}
the difference is greater than zero (except if $a=0$ that reduces to
Schwarzschild black hole). On this basis, one might conjecture that NLED
black holes are unstable.  However, above was shown that SSS BI solutions
are stable under linear perturbations. Hence for BI black hole there is no
relation between the positivity of the masses difference and stability
under perturbations of the metric and the electromagnetic potential.
 
ACS stability conjecture has been tested for other SSS NLED solutions (in
the magnetic sector), proving to be true; we conclude then that ACS
unstability conjecture does not apply generically for NLED solutions.

In relation to Born-Infeld black hole stability recently was presented in
\cite{Fernando} the analysis of quasinormal modes for the gravitational
perturbations, deriving a one dimensional Schrodinger type wave equation
for the axial perturbations. From the behavior of the potentials it was
concluded in \cite{Fernando} that the EBI black holes are classically
stable, this in agreement with the analysis based on Lagrangian
inequalities.
  
Current trends in Born-Infeld black holes include the coupling of EBI
fields with dilaton and axion field as well as non Abelian Born Infeld
structures \cite{Kunz}.  Another aspect that has been explored is BI black
holes as gravitational lenses: deflections depend on the BI parameter $b$
\cite{Eiroa}, \cite{Mosquera}. Lately also has been arised interest in BI
black hole thermodynamics in extra dimensions \cite{Riazi}.

%***********************


\begin{thebibliography}{99}

\bibitem{BI}
Born, M., Infeld, L.: {\it Foundations of the New Field Theory},
Proc. R. Soc. (London) {\bf A144}, 425-451 (1934).

\bibitem{hoffmann} 
Hoffmann, B., Infeld, L.: {\it On the Choice of the Action Function in the
New Field Theory}, Phys. Rev. {\bf 51} 765-773, (1937).

\bibitem{Peres}
Peres, A. :{\it Nonlinear Electrodynamics in General Relativity}
Phys. Rev. {\bf 122} 273-274, (1961).

\bibitem{PT}
Pellicer, R., Torrence, R. J.:{\it Nonlinear Electrodynamics and General
Relativity}, J. Math. Phys. {\bf 10}, 1718-1723 (1969).

\bibitem{Pleban1}
Pleba\~nski, J. F.:{\it Lectures on Non-linear Electrodynamics}, 
(Copenhagen, NORDITA, 1970).

\bibitem{Pleban2}
Salazar, H., Garc\'{\i}a, A., Pleba\~nski, J.F.: {\it Duality rotations
and type D solutions to Einstein equations with nonlinear electromagnetic
sources}, J. Math. Phys. {\bf 28}, 2171-2181, (1987).

\bibitem{GSP}
Garc\'{\i}a, A., Salazar, H., Pleba\~nski, J. F.: {\it Type-D solutions of
the Einstein and Born-Infeld Nonlinear Electrodynamics Equations}, Nuovo
Cim. {\bf 84}, 65-90 (1984).

\bibitem{Pleban3}
Dudley, A., Alarc\'on, S., Pleba\~nski, J. F.: {\it Signals and
discontinuities in general relativistic nonlinear electrodynamics}, J.
Math. Phys. {\bf 22}, 2835-2848 (1981).

\bibitem{Demian}
Demianski, M.: {\it Static Electromagnetic Geon}, Found. of Phys. {\bf
16}, 187-190 (1986).

\bibitem{HE} 
Heisenberg, W., Euler, H.: {\it Folgerungen aus der Diracschen Theorie des
Positrons}, (1936) Zeit. f. Phys. {\bf 98} 714-732. Weisskopf, W.S. {\it
On the self-energy and the electromagnetic field of the electron}, Phys.
Rev. {\bf 56} 72 (1939).
  
\bibitem{Schwinger} Schwinger, J.: {\it On Gauge Invariance and Vacuum
Polarization}, {Phys. Rev.}{\bf 82} 664-679 (1951).

\bibitem{Delphenich} 
Delphenich, D. H.:{\it Nonlinear Electrodynamics and QED},
arXiv: hep-th/0309108.

\bibitem{Gibbons} 
Gibbons, G. W.: {\it Born-Infeld particles and Dirichlet p-branes}, Nucl.
Phys. {\bf B 514} 603-639 (1998).

\bibitem{Ayon} 
Ay\'on-Beato, E., Garc\'{\i}a, A.: {\it Regular black hole in general
relativity coupled to nonlinear electrodynamics}, Phys. Rev. Lett., {\bf
80}, 5056-5059 (1998); Ay\'on-Beato, E., Garc\'{\i}a, A. :{\it
Nonsingular charged black hole solution for nonlinear source}, Gen. Rel.
Gravit., {\bf 31}, 629-633 (1999); Ay\'on-Beato, E., Garc\'{\i}a, A.: 
{\it New regular black hole solution from nonlinear electrodynamics},
Phys. Lett. B {\bf 464}, 25-28, (1999).

\bibitem{Bronnikov2} 
Bronnikov, K. A.:  {\it Comment on ``Regular black hole in general
relativity coupled to nonlinear electrodynamics"}, Phys. Rev. Lett. {\bf
85}, 4641 (2000).

\bibitem{Bronnikov}
Bronnikov, K. A.: {\it Regular magnetic black holes and monopoles from
nonlinear electrodynamics}, Phys. Rev. D {\bf 63}, 044005 (2001).

\bibitem{Novello}
Novello, M. De Lorenci, V. A., Salim, J. M., Klippert, R.: {\it
Geometrical Aspects of Light Propagation in Nonlinear Electrodynamics}
Phys. Rev. D. {\bf 61} (2000) 045001.

\bibitem{Novello2}
Novello, M., Perez Bergliaffa, S.E., Salim, J. M.: {\it Singularities in
general relativity coupled to nonlinear electrodynamics} Class. Quant.
Grav. {\bf 17}, 3821-3831, (2000).

\bibitem{Dymnikova} 
Dymnikova, I : {\it Regular electrically charged vacuum structures with de
Sitter center in Nonlinear Electrodynamics coupled to General Relativity
}, Class. Quant.  Grav. {\bf 21}, 4417-4429, (2004).

\bibitem{Gibbons-Rasheed} 
Gibbons, G. W., Rasheed, D. A.: {\it Electric-magnetic duality rotations
in non-linear electrodynamics}, Nucl. Phys. {\bf B 454} 185 (1995).

\bibitem{PP}
Pleba\~nski, J. F., Przanowski, M.: {\it Duality Transformations in
Electrodynamics}, Int. J. Theor. Phys. {\bf 33} 1535-1551, (1994).

\bibitem{Gibbons-Hashimoto}
Gibbons, G. W., Hashimoto, K.: {\it Non-linear Electrodynamics in Curved
Backgrounds}, JHEP {\bf 0009} 013 (2000).
 

\bibitem{Graves} 
Graves, J. C., Brill D. R.: {\it Oscillatory Character of
Reissner-Nordstr\"om Metric for an Ideal Charged Wormhole }, Phys. Rev.
{\bf 120} 1507-1513 (1960)

\bibitem{Breton1}
Bret\'on, N.: {\it Geodesic structure of the Born-Infeld black hole},
Class. Quant. Grav. {\bf 19}, 601-612, (2002).

\bibitem{Boillat}
Boillat, G.: {\it Nonlinear Electrodynamics: Lagrangians and Equations of
Motion}, J. Math. Phys. {\bf 11} 941-951 (1970).

%%%%%%%%%%%%%%%%%%%%%%%%%%%%%%%%


\bibitem{Wald}
Wald, R.: {\it The First Law of Black Hole Mechanics},
In `College Park 1993, Directions in general relativity, vol. 1'
358-366. arXiv: gr-qc/9305022.

\bibitem{ACS}
Ashtekar, A., Corichi, A., Sudarsky, D.: {\it Hairy black holes,
horizon mass and solitons}, Class.Quant.Grav. {\bf 18},
919-940, (2001).

\bibitem{Heusler}
Heusler, M., Straumann, N.: {\it The Fist law of black hole physics for a
class of nonlinear matter models}, Class. Quant. Grav. {\bf 10},
1299-1322, (1993).

\bibitem{Rasheed}
Rasheed, D. A.: {\it Non-Linear Electrodynamics: Zeroth and First
Laws of Black Hole Mechanics}, arXiv: hep-th/9702087.

\bibitem{ulises}
Corichi, A., Nucamendi, U., Sudarsky, D.: {\it Einstein-Yang-Mills
isolated horizons: phase space, mechanics, hair and conjectures}, Phys.
Rev. {\bf D 62}, 044046 (2000).

\bibitem{ABGBardeen}
 Ay\'on-Beato, E., Garc\'{\i}a, A.: {\it The Bardeen model as a nolinear
magnetic monopole}, Phys. Lett. B {\bf 493}, 149-152, (2000).

%%%%%%%%%%%%%%%%%%%%%%%%%%%%%%%%%%%%%

\bibitem{Ashtekar} 
Ashtekar, A., Fairhurst, S., Krishnan, B.: {\it Isolated horizons:
Hamiltonian evolution and the first law}, Phys. Rev. {\bf D 62},
104025 (2000).
 
\bibitem{Breton2} 
Bret\'on, N.: {\it  Born-Infeld black hole in the isolated horizon
framework}, Phys. Rev. {\bf D 67}, 124004 (2003).

\bibitem{M-S}
Moreno, C., Sarbach, O. {\it Stability properties of black holes in
selfgravitating nonlinear electrodynamics},
Phys. Rev. D {\bf 67}, 024028
(2003).

\bibitem{Breton3}
Bret\'on, N.: {\it  Stability of nonlinear magnetic black holes
}, Phys. Rev. {\bf D 72}, 044015 (2005).

\bibitem{Fernando}
Fernando, S.: {\it Gravitational perturbations and qua\-si-nor\-mal
modes of charged black holes in Eins\-tein-Born-In\-feld gravity},
Gen. Rel. Grav. {\bf 37} 585-604 (2005). hep-th/0407062.

\bibitem{Kunz}
Wirschins, M., Sood, A., Kunz, J.: {\it Non-Abelian Einstein-Born-Infeld
black holes}, Phys. Rev. {\bf D 63}, 084002 (2001).

\bibitem{Eiroa}
Eiroa, E. F.: {\it Gravitational lensing by Einstein Born Infeld Black
holes}, Phys. Rev. {\bf D 73}, 043002 (2006).

\bibitem{Mosquera}
Mosquera-Cuesta, H. J., de Freitas Pacheco, J. A., Salim, J. A.: {\it
Einstein's gravitational lensing and nonlinear electrodynamics}, Int. J.
Mod. Phys. {\bf A 21}, 43-55 (2006).


\bibitem{Riazi}
Sheykhi, A., Riazi, N.: {\it thermodynamics of black holes in (n+1)
dimensional Einstein-Born-Infeld dilaton gravity} arXiv: hep-th/0610085. 


\end{thebibliography}
\end{document}